\documentclass[letterpaper, 10 pt, conference]{ieeeconf}  

\IEEEoverridecommandlockouts                              

\usepackage{graphicx} 

\usepackage{lipsum}
\usepackage{xcolor}
\usepackage[binary-units]{siunitx}
\usepackage{booktabs}
\usepackage{tikz}
\usepackage{atbegshi}
\usepackage{pgfplots} 
\usepackage{multirow}

\newcommand{\fig}[1]{Fig.~\ref{#1}}

\newcommand{\tab}[1]{Table~\ref{#1}}

\newcommand{\sect}[1]{Section~\ref{#1}}

\newcommand\copyrighttext{%
    \footnotesize \copyright{ }2024 IEEE. Personal use of this material is permitted. Permission from IEEE must be obtained for all other uses, in any current or future media, including reprinting/republishing this material for advertising or promotional purposes, creating new collective works, for resale or redistribution to servers or lists, or reuse of any copyrighted component of this work in other works.}
\newcommand\copyrightnotice{%
    \AtBeginShipoutNext{\AtBeginShipoutUpperLeftForeground{
        \begin{tikzpicture}[remember picture,overlay]
            \node[anchor=south,yshift=15pt,xshift=0pt] at (current page.south) {\parbox{\dimexpr\textwidth-\fboxsep-\fboxrule\relax}{\copyrighttext}};
        \end{tikzpicture}%
    }}
}

\newcommand\httpsurl[1]{%
  \href{https://#1}{\nolinkurl{#1}}%
}

\makeatletter
\let\NAT@parse\undefined
\makeatother
\usepackage{hyperref}

\title{\LARGE \bf
V2AIX: A Multi-Modal Real-World Dataset of \\ ETSI ITS V2X Messages in Public Road Traffic
}

\author{Guido Kueppers$^{\dagger}$, Jean-Pierre Busch$^{\dagger}$, and Lennart Reiher$^{\dagger}$, Lutz Eckstein$^{\ddagger}$
\thanks{$^{\dagger}$The authors contributed equally to this work. They are with the Institute for Automotive Engineering~(ika), RWTH Aachen University, Germany. {\tt\small \{firstname.lastname\}@ika.rwth-aachen.de}}
\thanks{$^{\ddagger}$Lutz Eckstein is head of the Institute for Automotive Engineering~(ika).}
}

\begin{document}

\maketitle

\thispagestyle{empty}
\pagestyle{empty}

\copyrightnotice


\begin{abstract}

Connectivity is a main driver for the ongoing megatrend of automated mobility: future Cooperative Intelligent Transport Systems~(C-ITS) will connect road vehicles, traffic signals, 
roadside infrastructure, and even vulnerable road users, sharing data and compute for safer, more efficient, and more comfortable mobility.
In terms of communication technology for realizing such vehicle-to-everything~(V2X) communication, the WLAN-based peer-to-peer approach (IEEE 802.11p, ITS-G5 in Europe) competes with C-V2X based on cellular technologies (4G and beyond).
Irrespective of the underlying communication standard, common message interfaces are crucial for a common understanding between vehicles, especially from different manufacturers.
Targeting this issue, the European Telecommunications Standards Institute~(ETSI) has been standardizing V2X message formats such as the Cooperative Awareness Message~(CAM).
In this work, we present \textit{V2AIX}, a multi-modal real-world dataset of ETSI ITS messages gathered in public road traffic, the first of its kind.
Collected in measurement drives and with stationary infrastructure, we have recorded more than \num{285000} V2X messages from more than \num{2380} vehicles and roadside units in public road traffic.
Alongside a first analysis of the dataset, we present a way of integrating ETSI ITS V2X messages into the Robot Operating System~(ROS).
This enables researchers to not only thoroughly analyze real-world V2X data, but to also study and implement standardized V2X messages in ROS-based automated driving applications.
The full dataset is publicly available for non-commercial use at \httpsurl{v2aix.ika.rwth-aachen.de}.

\end{abstract}

\section{Introduction}

\label{sec:introduction}

Connectivity is the key driver behind realizing cyber-physical systems~(CPS) and an Internet-of-Things~(IoT) in the age of Industry~4.0. Connectivity also takes automated driving to the next level: vehicles, traffic signals, and roadside infrastructure communicate with control centers and other traffic participants to form large-scale Cooperative Intelligent Transport Systems~(C-ITS). Such connectivity between so-called ITS~stations is collectively known as vehicle-to-everything~(V2X) communication, including, e.g., vehicle-to-vehicle~(V2V), vehicle-to-infrastructure~(V2I), vehicle-to-person~(V2P), but also the other way around.

With WLAN- and cellular-based V2X~(C-V2X), there are currently two competing technical approaches prevalent for V2X communication. WLAN-based dedicated short-range communications~(DSRC) is built on top of the IEEE~802.11p WLAN standard, enabling direct peer-to-peer communication between nearby traffic participants. In Europe, this system is implemented as ITS-G5, standardized by the European Telecommunications Standards Institute~(ETSI). In contrast, C-V2X, as standardized by 3GPP, is built on top of 4G~LTE or 5G mobile cellular connectivity. C-V2X allows for two interfaces, a wide area network interface~(Uu) and a direct peer-to-peer communications interface~(PC5). While both ITS-G5 and C-V2X are present in research and industry, adoption in North America and China trends towards C-V2X~\cite{iot2024cv2x}. In 2019, the EU has announced to adopt a technology-neutral approach to C-ITS~\cite{capacity2019eu}.

As of 2024, V2X technology in Europe remains divided: the \textit{Volkswagen Group} has been rolling out ITS-G5 in series since 2020, other manufacturers like \textit{BMW} and \textit{Mercedes-Benz} offer C-V2X-based (non-peer-to-peer) V2X features as a premium option~\cite{ADAC2024v2x}.


\begin{figure}[t]
    \centering
    \smallskip
    \smallskip
    \includegraphics[width=1\linewidth]{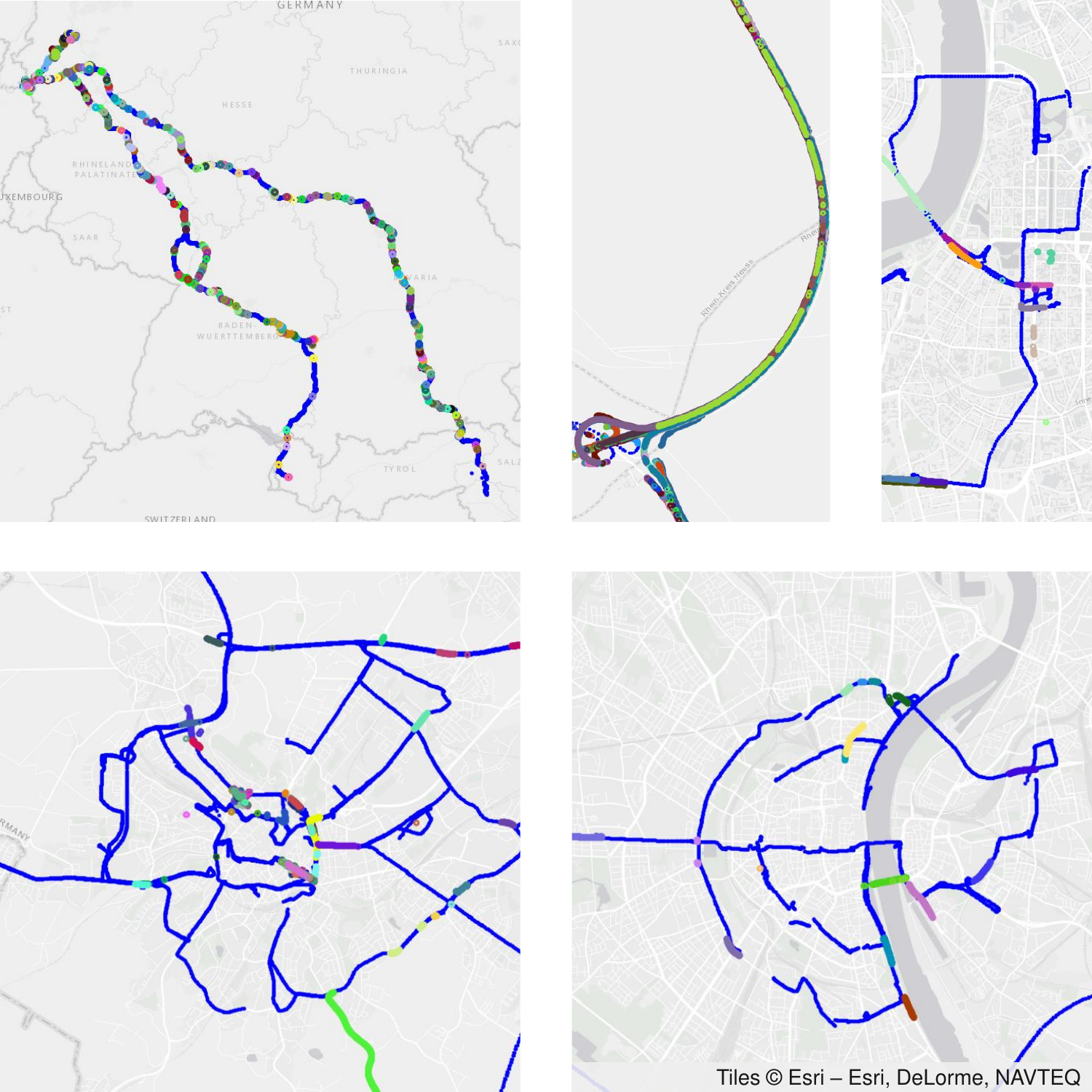}
    \caption{V2X messages at core recording locations collected in the \textit{V2AIX} dataset: driven routes in blue, reference positions of received V2X messages colored by unique ITS station. From top-left to bottom-right: overview, A44, Dusseldorf, Aachen (\textit{Aix-la-Chapelle}) and Cologne.}
    \label{fig:teaser}
\end{figure}


Irrespective of the underlying communication technology, common message interfaces are required for a common understanding between vehicles and infrastructure, especially across manufacturers and other service providers. In this regard, ETSI has not only been standardizing the \mbox{ITS-G5} technology, but also application-layer message formats: CAMs (Cooperative Awareness Messages) transmit status information such as position of the originating ITS stations; DENMs (Decentralized Environmental Notification Messages) alert road users of events such as emergency braking; SPATEMs (Signal Phase and Timing Extended Messages) in conjunction with MAPEMs (Map Topology Extended Messages) provide information about signal changes, e.g., at intersections; although not fully-standardized yet, more message types for V2X use cases are in the works, e.g., CPM for collective perception or MCM for maneuver coordination.

Although these standardized V2X messages are already exchanged in real-world road traffic, we find that they are not yet present in any large-scale dataset for automated driving research and development, see \sect{sec:v2x-datasets}. We hope to start filling this gap with the first release of the \textit{V2AIX} dataset. The collected ETSI ITS V2X messages offer valuable insights into the current status of standardized V2X in public road traffic: how often does one encounter V2X-capable vehicles; to what extent are the generic message types filled in reality; do the transmitters adhere to the standard?

In addition to offering the potential for such data analyses, the \textit{V2AIX} dataset also makes real-world V2X messages more accessible. Alongside the dataset itself, we release a method of integrating ETSI ITS messages into the Robot Operating System~(ROS): the \texttt{etsi\_its\_messages} package stack. Using this integration, researchers and engineers are presented with a smaller barrier to incorporate standardized V2X messaging into prototypical automated driving applications.

\subsection{Contributions}

In summary, the contribution of this work is three-fold:

\begin{itemize}
    \item we collect and make publicly available the novel dataset \textit{V2AIX}: a multi-modal real-world dataset of ETSI ITS V2X messages in public road traffic;
    \item we screen, analyze, and discuss the recorded ETSI ITS messages and suggest use cases for further usage of the data;
    \item we release the open-source ROS / ROS~2 package stack \texttt{etsi\_its\_messages}\footnote{\httpsurl{github.com/ika-rwth-aachen/etsi_its_messages}}, greatly facilitating the usage of ETSI ITS messages in ROS as well as across the borders of the ROS ecosystem.
\end{itemize}

\section{Related Work}
\label{sec:related-work}

Driven by a growing interest and significant investments in automated driving over the past decade, both public as well as private actors have been publishing openly accessible datasets for automated driving. The public release of these datasets has enabled researchers worldwide to study and evaluate novel approaches for automated driving-related tasks on real-world (and simulated) data.~\cite{LiuEtAl_SurveyAutonomousDriving_2024}

\subsection{Datasets for Automated Driving}

With one of the most recent and most complete surveys, the authors of~\cite{LiuEtAl_SurveyAutonomousDriving_2024} have sighted and analyzed over 200 datasets for automated driving. Alongside an impact assessment, the survey classifies datasets in terms of sensors and modalities, autonomous driving tasks, and annotation processes.

A large portion of available datasets is targeted at perception tasks, i.e., semantic segmentation, object detection, object tracking, and more. In chronological order, the most influential perception datasets according to~\cite{LiuEtAl_SurveyAutonomousDriving_2024} include \textit{KITTI}~\cite{GeigerEtAl_VisionMeetsRobotics_2013}, \textit{CityScapes}~\cite{CordtsEtAl_CityscapesDatasetSemantic_2016a}, \textit{nuScenes}~\cite{CaesarEtAl_NuScenesMultimodalDataset_2020}, \textit{Waymo Open Dataset}~\cite{SunEtAl_ScalabilityPerceptionAutonomous_2020}, and \textit{Argoverse~2}~\cite{WilsonEtAl_ArgoverseNextGeneration_2023}.

As a result of precise object annotations, many perception datasets are automatically also suitable for the downstream tasks of trajectory prediction and trajectory planning. In addition, several task-specific datasets exist that have specifically collected trajectory data. While perception datasets mostly rely on onboard sensors of measurement vehicles for data collection, trajectory datasets have also been recorded using stationary roadside infrastructure sensors~\cite{LaurentKlokerEtAl_HowBuildHighly_2021, KrammerEtAl_ProvidentiaLargeScaleSensor_2022, YuEtAl_DAIRV2XLargeScaleDataset_2022a} or camera-equipped UAVs, i.e., drones. Notable datasets include \textit{NGSIM}~\cite{USDepartmentofTransportation:FederalHighwayAdministration_USHighway101_}, \textit{INTERACTION}~\cite{ZhanEtAl_INTERACTIONDatasetINTERnational_2019}, and the \textit{leveLXData} family of datasets~(\textit{highD}, \textit{inD}, \textit{exiD}, \textit{rounD}, \textit{uniD})~\cite{KrajewskiEtAl_HighDDatasetDrone_2018, BockEtAl_InDDatasetDrone_2020, KrajewskiEtAl_DatasetDroneDataset_2020, MoersEtAl_ExiDDatasetRealWorld_2022}.

\begin{figure*}[t]
    \centering
    \smallskip
    \includegraphics[width=\textwidth]{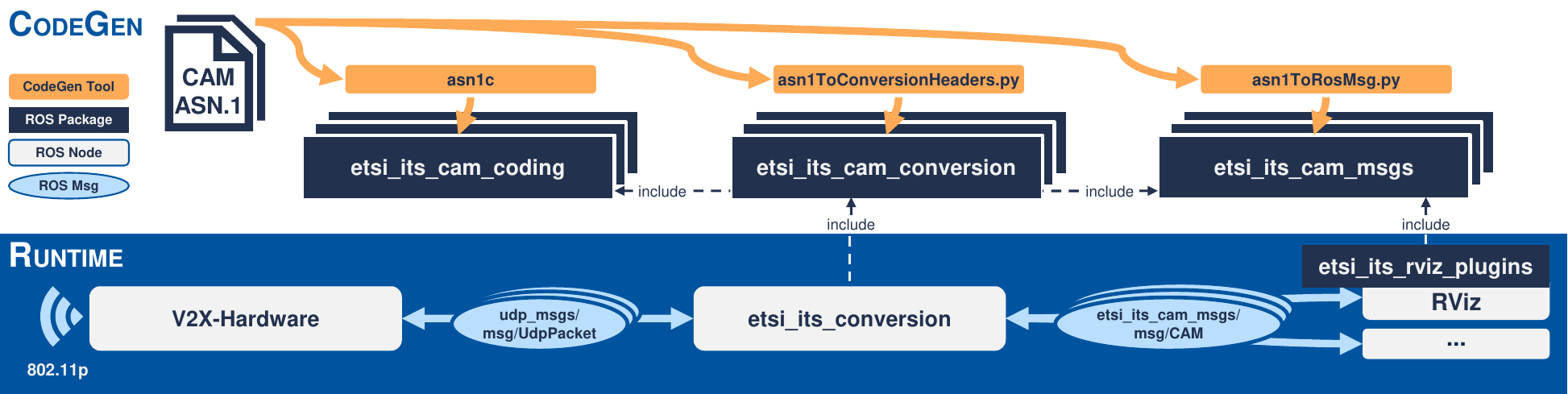}
    \caption{Concept of \texttt{etsi\_its\_messages}: 1) \textit{Code Generation}: based on ASN.1 definitions, C++ libraries containing C-struct implementations of ETSI ITS message types including encoding/decoding support are generated using \texttt{asn1c}~\cite{LevWalkin_asn1c} (\textit{coding}); ROS message type definitions are generated (\textit{msgs}); conversion functions between C-structs and ROS message objects are generated (\textit{conversion}); 2) \textit{Runtime}: V2X messages are received/transmitted with dedicated hardware, e.g., via ITS-G5, or simulated; a simple driver publishes UDP message payloads as ROS messages; the \texttt{etsi\_its\_conversion} node decodes the payloads and converts them to the ROS messsage equivalents of ETSI ITS message types; downstream ROS applications like RViz for visualization work with ROS V2X messages.}
    \label{fig:etsi2ros}
\end{figure*}

\subsection{V2X Datasets}
\label{sec:v2x-datasets}

In the context of automated driving datasets, a dedicated reference to V2X usually translates to having multiple views of the same scene, collected by sensors of multiple traffic participants~\cite{HuangEtAl_V2XCooperativePerception_2023}. The main purpose of such datasets therefore is to facilitate research in the area of cooperative perception. Interest in dedicated datasets for cooperative V2X tasks has recently been increasing~\cite{LiuEtAl_SurveyAutonomousDriving_2024}. Apart from multiple synthetic cooperative perception datasets~\cite{XuEtAl_OPV2VOpenBenchmark_2022, LiEtAl_V2XSimMultiAgentCollaborative_2022, XuEtAl_V2XViTVehicletoEverythingCooperative_2022}, \textit{V2V4Real}~(V2V)~\cite{XuEtAl_V2V4RealRealWorldLargeScale_2023}, \textit{DAIR-V2X}~(V2I)~\cite{YuEtAl_DAIRV2XLargeScaleDataset_2022a}, and \textit{LUCOOP}~(V2V, V2I)~\cite{AxmannEtAl_LUCOOPLeibnizUniversity_2023} are notable mentions for real-world V2X cooperative perception datasets.

When it comes to datasets containing actual V2X communication, e.g., in the form of standardized ETSI ITS messages, a large gap is identified. To the best of our knowledge, only two very specific datasets of real-world V2X communication in Europe have been published. None of the two datasets includes multi-modal sensor data for context information.

The authors of \cite{MavromatisEtAl_OperatingITSG5DSRC_2019} present a dataset of network interactions between two V2X on-board units~(OBUs) and four roadside units~(RSUs) along the FLOURISH test track in Bristol, UK. The dataset refers to eight experiments of two hours each, logging CAMs every \SI{10}{\milli\second}. As the authors focus on investigating V2X performance across different telecommunication bands, only CAMs self-generated by the two vehicles and four RSUs are considered. While PCAP~traces of the communication are provided, only selected quantities of the transmitted CAMs and GPS measurements are provided as plain CSV~files.

The authors of \cite{_V2XMeasurementM0_} publish a dataset of V2X communication between four OBU-equipped vehicles and multiple RSUs along \SI{72}{\kilo\meter} of the M0 motorway in Budapest, Hungary. The data provided in JSON format comprises of decoded V2X messages both transmitted (CAM, DENM) and received (CAM, DENM, MAPEM, SPATEM, IVIM) by the four measurement vehicles.

Outside of Europe, the US Department of Transportation's \textit{ITS DataHub}~\cite{_ITSDataHub_} has collected several V2X-related datasets from field tests across the US, also including V2X messages in the form of BSMs~(Basic Safety Messages), the US's equivalent to CAMs.

\subsection{ROS Tooling for ETSI ITS V2X Messages}
\label{sec:ros-tooling}

The \textit{Robot Operating System~(ROS)} is an open-source set of software libraries and tools for building robotic applications~\cite{quigley2009ros, macenski2022ros2}. This way, it has become one of the most widely used robotics frameworks and has penetrated both academia and industry. Especially in academia, it is also one of the go-to underlying frameworks for building automated driving applications and entire automated driving software stacks such as \textit{Autoware}~\cite{kato2018autoware}.

In the context of V2X and, more specifically, ETSI~ITS messaging, there exist open-source repositories that provide message equivalents for ROS~\cite{RaphaelRiebl_Ros_etsi_its_msgs_, DLRTransportationSystems_V2x_if_ros_msg_}. Other publications have implemented closed-source versions of ROS support for V2X messages~\cite{ShanEtAl_DemonstrationsCooperativePerception_2020, YedidaEtAl_V2XConnectivityROS_2021, EliasEtAl_EmergingV2XParadigm_2022}.

Going beyond ROS message equivalents for ETSI~ITS message types, the \texttt{vehicleCAPTAIN} toolbox~\cite{ChristophPilz_Vehicle_captain_toolbox_} provides auto-generated ROS~2 messages as well as a gateway for translating ASN.1-encoded ETSI ITS messages to their ROS counterparts and vice-versa.

A main component of our dataset collection pipeline is our newly open-sourced \texttt{etsi\_its\_messages} ROS package stack, following a similar approach than the \texttt{vehicleCAPTAIN} toolbox, cf.~\sect{sec:etsi2ros}.

 At the time of our release, the \texttt{vehicleCAPTAIN} toolbox supports more ETSI ITS message types than our \texttt{etsi\_its\_messages} stack, but the conversion architecture is more complex, there is no backwards compatibility for ROS~1, and it is not officially released as a ROS package.

\section{ROS-Integration for ETSI ITS V2X Messages}
\label{sec:etsi2ros}

\begin{figure*}[t]
    \centering
    \smallskip
    \includegraphics[width=\textwidth]{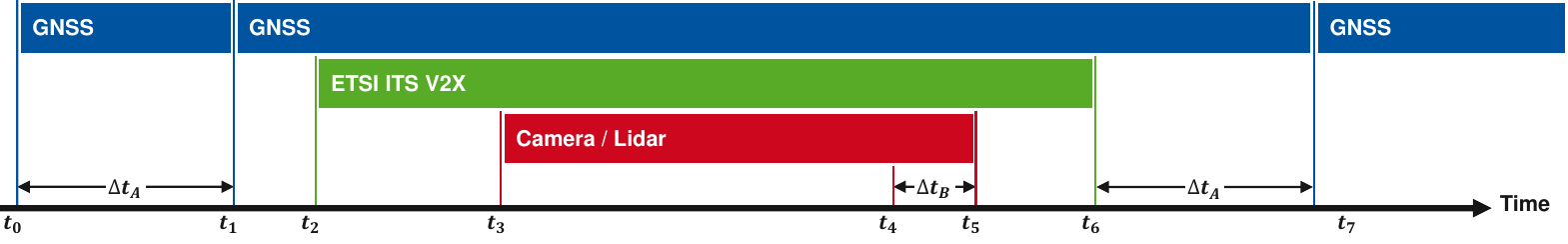}
    \caption{Automated data collection procedure: \( t_0 \)) GNSS data is recorded to a new ROS bag file; \( t_1 \)) no ETSI ITS V2X message has been received for \( \Delta t_A = t_1 - t_0 \), recording to a new ROS bag file is started; \( t_2 \)) ETSI ITS V2X messages are received and recorded; \( t_3 \)) reference position of received CAM is within \( d_{\mathrm{min}} \), camera images and lidar point clouds are recorded; \( t_4 \)) received CAM is beyond \( d_{\mathrm{min}} \); \( t_5 \)) no received CAM has been within \( d_{\mathrm{min}} \) for \( \Delta t_B = t_5 - t_4 \), recording of camera and lidar data is stopped; \( t_6 \)) last ETSI ITS V2X message for a while is received; \( t_7 \)) no ETSI ITS V2X message has been received for \( \Delta t_A = t_7 - t_6 \), recording to a new ROS bag file is started. A single bag file may contain multiple sequences of V2X messages and sensor data.}
    \label{fig:data-collection}
\end{figure*}

As emphasized in~\sect{sec:ros-tooling}, within the context of automated driving, a key component missing in ROS is dedicated support for V2X messaging. Although ROS supports the definition of arbitrary custom message types -- a feature widely used for propagating data through an automated driving stack -- communication between vehicles and infrastructure, also going across the border of individual ROS systems, is not properly supported. We aim at filling this gap with our open-source release of \texttt{etsi\_its\_messages}: a set of packages for supporting the standardized ETSI ITS V2X messages in ROS and ROS~2. The ROS integration is two-fold:
\begin{enumerate}
    \item the definition of nested ROS message types equivalent to the standardized ETSI ITS message types such as CAM, DENM and CPM;
    \item and the implementation of an application for bi-directionally bridging into and out of the ROS ecosystem, making it possible to connect a ROS-based stack to V2X messaging outside of ROS.
\end{enumerate}

The core concept of \texttt{etsi\_its\_messages} is to automatically generate the ROS support code based on the standardized ASN.1 definitions of the ETSI ITS messages~\cite{ETSI2024gitlab}. As the standardized ETSI ITS message types are complex and deeply nested, automated code generation is the only feasible method for maintaining support for many ETSI ITS message types. To ease access of specific fields in these nested messages, we also provide useful access functions. See \fig{fig:etsi2ros} for more details on the code generation process.

In addition to the definition of ROS message equivalents, the provided ROS node \texttt{etsi\_its\_conversion} converts between ASN.1-encoded payloads and their ROS equivalents through automatically generated conversion functions. This allows to receive and send V2X messages from and to components outside of the ROS ecosystem. As part of our dataset collection pipeline~(cf.~\sect{sec:data-collection}), we receive such \mbox{ASN.1-encoded} V2X messages with a dedicated V2X modem and bridge these messages to ROS using the presented conversion node. See \fig{fig:etsi2ros} for more details on the runtime process.

Note that our ROS integration not only facilitates working with V2X messages in ROS alone, but also connects ROS-based systems to other ROS- or non-ROS-based systems. The external interface is the standardized ASN.1 encoding of ETSI ITS messages, also implemented by dedicated V2X hardware, simulation platforms like \textit{Artery}~\cite{riebl2015artery}, or open-source implementations of the ETSI ITS protocol suite such as \textit{Vanetza}~\cite{riebl2017vanetza} or \textit{OpenC2X}~\cite{laux2016openc2x}.

\section{V2AIX Dataset Collection}
\label{v2aix}

Utilizing the previously presented \texttt{etsi\_its\_messages} ROS package stack, we have collected the \textit{V2AIX} dataset: a multi-modal real-world dataset of ETSI ITS V2X messages in public road traffic, the first of its kind.

The first release of \textit{V2AIX} covers data recorded in urban and highway traffic around Aachen, Germany. Data is collected by test vehicles in measurement drives as well as by stationary infrastructure.

\subsection{Hardware Setup}

For mobile measurement drives, we use both our own sensor-equipped research vehicle for automated driving, as well as other V2X-only vehicles. The research vehicle carries three front-facing and one rear-facing camera (\textit{Stereolabs ZED 2i 4mm}), one front-facing FMCW lidar with velocity information (\textit{Aeva Aeris II}), one centered rotating lidar (\textit{Ouster OS-1}), and one rear-facing solid state lidar (\textit{Velodyne Velarray M1600EX}). The cameras and lidars are used to provide context information whenever V2X messages are recorded. For localization, we use a high-precision GNSS/INS unit (\textit{OXTS RT3000}). ITS-G5 V2X messages are received with a \textit{Cohda Wireless MK5 OBU}. The V2X-only vehicles are only equipped with the V2X OBU that is also providing GNSS localization in that case.

For stationary infrastructure measurements, we use both a sensor-equipped research infrastructure station as well as V2X-only deployments. The infrastructure station carries two tilted rotating lidars (\textit{Ouster OS-1}) and two cameras (\textit{Bosch FLEXIDOME}) at a height of approximately \SI{6}{\meter}. ITS-G5 V2X messages are received with a \textit{Cohda Wireless MK5 RSU} that is also providing GNSS localization. The V2X-only deployments only consist of the V2X RSU, locally installed at specific locations.

\subsection{Recording Locations}

For mobile measurement drives using our sensor-equipped research vehicle, we choose to cover urban and highway sections in the Aachen -- Cologne -- Dusseldorf triangle in Germany. In total, we cover a distance of \SI{310}{\kilo\meter} over a time span of \SI{9.6}{\hour} in urban areas. Additionally, a V2X-only vehicle covers rides from Aachen to Austria and back resulting in a total distance of \SI{2906}{\kilo\meter} for highway drives. Recording drives are spread over multiple days in January~2024. Most of the driving routes are shown in \fig{fig:teaser}.

For stationary infrastructure measurements, we employ one of the infrastructure stations of the \textit{ACCorD} test field~\cite{LaurentKlokerEtAl_HowBuildHighly_2021} at the A44 interchange Jackerath between Aachen and Dusseldorf, Germany. In addition, V2X-only setups are deployed to multiple locations in Aachen. Recordings are spread over multiple days in January~2024.

\subsection{Data Collection}
\label{sec:data-collection}

\begin{table*}[!t]
    \centering
    \smallskip
    \smallskip
    \caption{Key statistics of \textit{V2AIX}}
    \begin{tabular}{l r r r r r r r r r r}
        \toprule
         & \multicolumn{4}{c}{\textbf{\# Messages}} & \textbf{\# Unique} & \multicolumn{2}{c}{\textbf{Distance [\SI{}{\kilo\meter}]}} & \multicolumn{3}{c}{\textbf{Duration [\SI{}{\hour}]}} \\
        \textbf{Type of Measurement} & \textbf{CAM} & \textbf{DENM} & \textbf{MAPEM} & \textbf{SPATEM} & \textbf{ITS Stations} & \textbf{Ego} & \textbf{CAM} & \textbf{Total} & \textbf{V2X} & \textbf{Context} \\
        \midrule
        Mobile (urban) & \num{8512} & \num{23} & \num{2315} & \num{18192} & \num{72} & \num{309.80} & \num{52.87} & \num{9.59} & \num{1.48} & \num{0.57} \\
        Mobile (highway) & \num{25126} & \num{425} & \num{0} & \num{0} & \num{423} & \num{2905.71} & \num{282.70} & \num{29.80} & \num{2.90} & \num{0.04} \\
        Stationary (urban) & \num{90147} & \num{890} & \num{0} & \num{0} & \num{1529} & - & \num{275.79} & \num{704.56} & \num{18.29} & - \\
        Stationary (highway) & \num{139682} & \num{0} & \num{0} & \num{0} & \num{364} & - & \num{1377.01} & \num{17.02} & \num{7.40} & \num{1.03} \\
        \midrule
        \( \mathbf{\Sigma} \) & \textbf{\num{263467}} & \textbf{1338} & \textbf{\num{2315}} & \textbf{\num{18192}} & \textbf{\num{2388}} & \textbf{\num{3215.51}} & \textbf{\num{1988.37}} & \textbf{\num{760.97}} & \textbf{\num{30.07}} & \textbf{\num{1.64}} \\
        \bottomrule
    \end{tabular}
    \label{tab:key-facts}
\end{table*}

We record all data to ROS bag files. The recorded messages can be categorized into three types: 1)~localization, 2)~ETSI ITS V2X messages, and 3)~sensor data for context. By means of an automated recording pipeline, we make sure to store our position along all recordings, to save all received V2X messages, but to only record context sensor data when meaningful. See \fig{fig:data-collection} for more details on the automated data collection procedure.

\subsubsection{Localization}

We record latitude/longitude positions measured by the V2X OBU's or RSU's GNSS at \SI{6}{\hertz}. On drives with the sensor-equipped research vehicle, we additionally record higher-precision location using the on-board GNSS/INS unit at \SI{100}{\hertz}. Position data is also recorded during periods of no V2X reception.

\subsubsection{ETSI ITS V2X Messages}

The V2X modems are configured to forward all V2X traffic received via ITS-G5 to a UDP socket on the recording computer. We use open-source ROS tools to wrap the UDP packets in ROS messages that are recorded. The raw V2X payload is also run through the \texttt{etsi\_its\_conversion} node to yield human-readable ROS V2X messages that are recorded as well. This allows us to, e.g., easily implement the distance-based trigger of sensor data as detailed at \( t_3 \) in \fig{fig:data-collection}.

\subsubsection{Sensor Data for Context}

Camera images and lidar point clouds are recorded only if a received CAM's reference position is within \( d_{\mathrm{min}} = \SI{125}{\meter} \) of our research vehicle or within \( d_{\mathrm{min}} = \SI{300}{\meter} \) of our infrastructure sensor station. Cameras record images at \SI{10}{\hertz}. Lidars record point clouds at \SI{10}{\hertz} as well. Disregarding sensor built-in object detections of the FMCW lidar, which are included in the dataset, the sensor data is not annotated but only collected for context.

\subsection{Postprocessing \& Format}

The recorded ROS bag files are trimmed to sections where V2X messages have been received. This yields separate V2X scenario recordings with a leading and a trailing \SI{1}{\second}. In a parallel postprocessing step, all bag files of a single measurement drive or setup are joined, excluding large camera and lidar data. This is helpful to retrace the driven route, even when no V2X messages were recorded. Camera images are anonymized to comply with EU GDPR regulations.

The dataset is delivered in two formats:
\begin{enumerate}
    \item ROS / ROS 2 bag files allow to directly replay and visualize the data in ROS (\SI{1300}{GB});
    \item human-readable JSON files of GNSS and V2X messages simplify data analysis and processing, independent of ROS (\SI{21}{GB}).
\end{enumerate}

\section{Dataset Statistics and Evaluation}
\label{sec:evaluation}

In the following section, we take a closer look at the recorded dataset. We denote key statistics, present insights into the V2X message contents, and suggest use cases for further exploration of the data.

\subsection{Overview}
\label{sec:overview}

In order to gain an overview of the \textit{V2AIX} dataset, key statistics are presented in \tab{tab:key-facts}. In total, we have collected \num{285312} ETSI ITS messages, covering five different message types: CAM, DENM, MAPEM, and SPATEM. The CAMs represent \SI{1988}{\kilo\meter} of trajectory data from \num{2388} different vehicles and other ITS stations. Out of a total recording time of \SI{760.97}{\hour}, we have collected ETSI ITS messages during \SI{30.07}{\hour} as well as \SI{1.64}{\hour} of context sensor data (lidar and camera).

\subsection{Message Contents}

CAMs are designed to communicate relevant state information, e.g., position, of ITS stations. As such, they hold some of the most valuable information to exchange and are naturally the most often encountered message type. A CAM consists of four containers: the \textit{basic container}, the \textit{high-frequency container}, the \textit{low-frequency container} and the \textit{special vehicle container}. Based on samples from the dataset, we notice that the optional \textit{special vehicle container} is never filled. Basic state information available in the other containers is always set, except for some optional fields like vertical acceleration.

Alongside various CAMs, \textit{V2AIX} contains  \num{1338} DENMs that refer to \num{37} events in total. Looking at the transmitted cause information in \tab{tab:denm_stats}, traffic conditions, stationary vehicles and dangerous situations are reported. As an example, one vehicle leaving a highway for a rest area reports a dangerous situation with active intervention of the Automatic Emergency Braking~(AEB).

Furthermore, the \textit{V2AIX} dataset contains SPATEMs and MAPEMs from various intersections. The respective messages, which provide information about the intersection topology and the signal states of the associated traffic lights, were captured at intersections of the ACCorD~\cite{LaurentKlokerEtAl_HowBuildHighly_2021} test field in Aachen.

\begin{table}[!h]
    \centering
    \caption{Event types transmitted in DENMs}
    \begin{tabular*}{\linewidth}{llrr}
    \toprule
         \multicolumn{2}{c}{\textbf{\textit{eventType}}} &  & \\
         \textbf{\textit{causeCode}} & \textbf{\textit{subCauseCode}} & \textbf{\# Msgs} & \textbf{\# ITS Stat.} \\
         \midrule
         Traffic Condition (\num{1}) & Unavailable (\num{0}) & \num{688} & \num{30} \\ 
         Stationary vehicle (\num{94}) & Unavailable (\num{0}) & \num{391} & \num{4} \\
         Dang. situation (\num{99}) & Unavailable (\num{0}) & \num{218} & \num{1} \\
         Dang. situation (\num{99}) & Emerg. brake (\num{1}) & \num{5} & \num{1} \\
         Dang. situation (\num{99}) & AEB activated (\num{5}) & \num{2} & \num{1} \\
         \bottomrule
    \end{tabular*}
    \label{tab:denm_stats}
\end{table}

\subsection{ITS-G5-Enabled Vehicles}

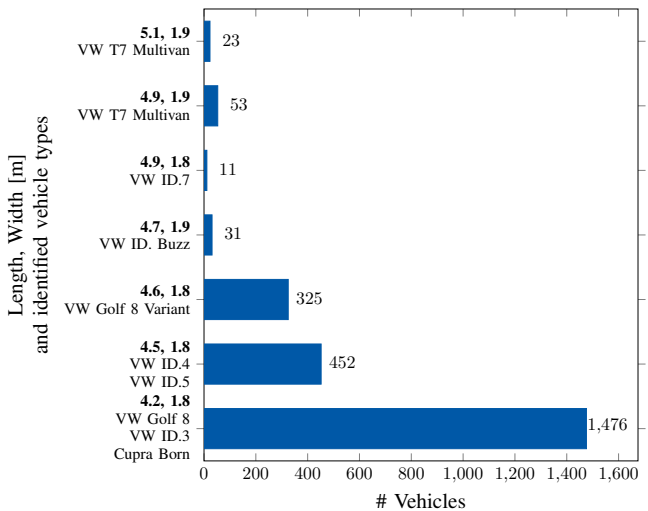
\begin{figure}[t]
    \centering
    \smallskip
    \smallskip

    \resizebox{\linewidth}{!}{%
        \begin{tikzpicture}
            \begin{axis} [xbar,
                        scale only axis,
                        height=9cm,
                        width=\linewidth,
                        bar width=0.8cm,
                        xlabel={\# Vehicles},
                        xlabel style={align=center,font=\large},
                        ylabel near ticks,
                        ylabel style={align=center,font=\large},
                        ylabel={Length, Width [m] \\ and identified vehicle types},
                        xticklabel style={font=\normalsize},
                        ytick={1,2,3,4,5,6,7},
                        yticklabel style={align=right, font=\small},
                        yticklabels={{\textbf{4.2, 1.8}\\VW Golf 8\\VW ID.3\\Cupra Born},{\textbf{4.5, 1.8}\\VW ID.4\\VW ID.5},{\textbf{4.6, 1.8}\\VW Golf 8 Variant},{\textbf{4.7, 1.9}\\VW ID. Buzz},{\textbf{4.9, 1.8}\\VW ID.7},{\textbf{4.9, 1.9}\\VW T7 Multivan},{\textbf{5.1, 1.9}\\VW T7 Multivan}}, 
                        ymin=0.5, ymax=7.5,
                        xmin=0, xmax=1675,
                        nodes near coords, 
                        nodes near coords align={vertical}, 
                        every node near coord/.append style={xshift=12pt, yshift=-6pt, font=\normalsize},
                        nodes near coords style={black}, 
                        ]
                \addplot[draw={rgb:green, 84;blue, 159},fill={rgb:green, 84;blue, 159}] coordinates {
                    (1476,1) 
                    (452,2) 
                    (325,3) 
                    (31,4)
                    (11,5)
                    (53,6)
                    (23,7)
                };
            \end{axis}
        \end{tikzpicture}
    }
    \caption{Number of unique vehicles per object dimension identified in all CAMs}
    \label{fig:vehdims}
\end{figure}

The \textit{V2AIX} dataset offers the possibility to analyze the penetration of ITS-G5 into production vehicles. Although CAM messages are not designed to carry information about the vehicle model, conclusions can be drawn from evaluating the encountered vehicle dimensions. We have captured CAMs containing seven unique pairs of vehicle dimensions. According to~\cite{ADAC2024v2x}, only \textit{Volkswagen Group} vehicles are equipped with ITS-G5 technology at the moment. Combining context images and vehicle dimensions, we can match specific vehicle models to specific CAM trajectories. As shown in \fig{fig:vehdims}, the compact \textit{Volkswagen} models \textit{ID.3} (or the similar \textit{Cupra Born}) and \textit{Golf 8} are most common. Note that some drivers may have deactivated the V2X feature.

\subsection{Qualitative Analysis of Localization Accuracy}
\label{sec:campos}

\begin{figure}
    \centering
    \includegraphics[width=\linewidth]{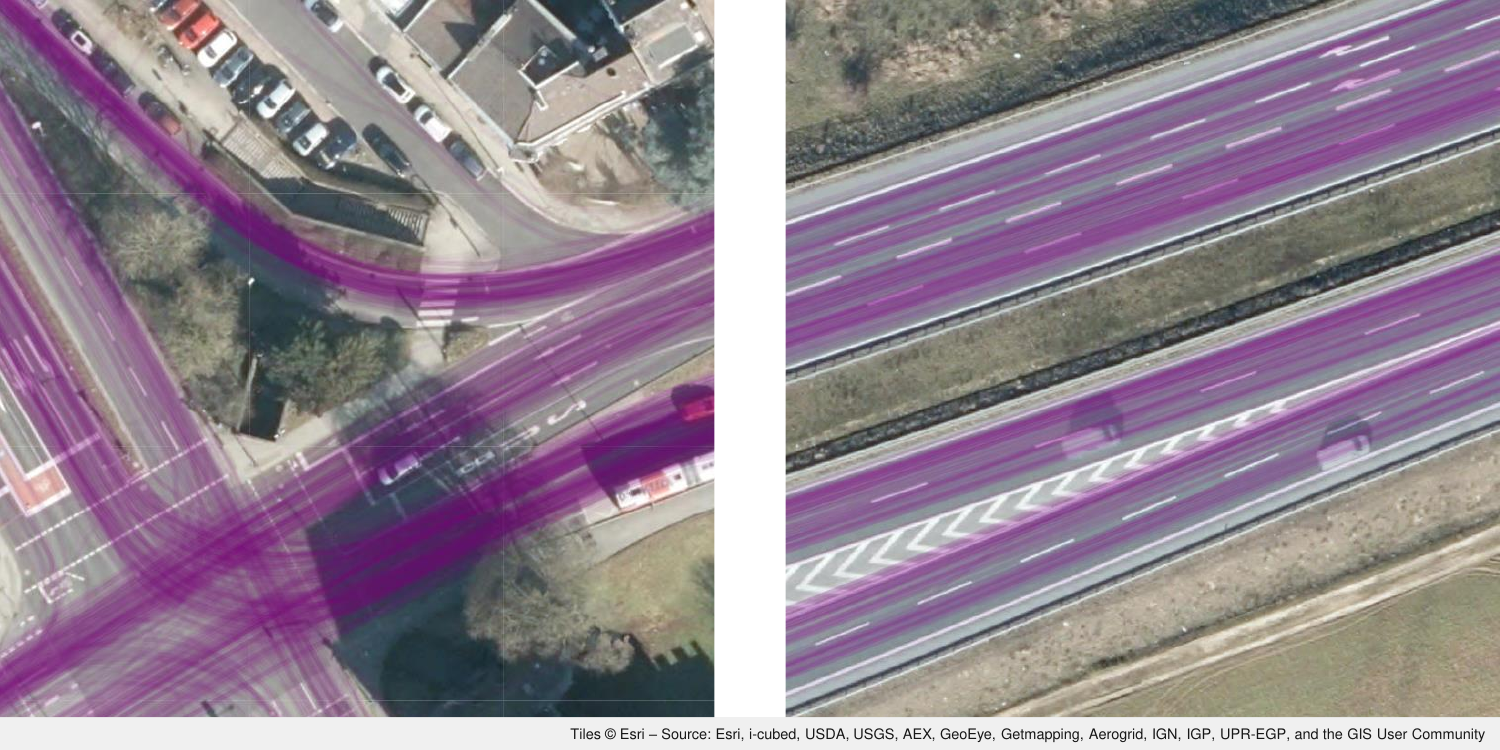}
    \caption{CAM trajectories collected by an urban (left) and a highway (right) stationary infrastructure station}
    \label{fig:posanalysis}
\end{figure}

\fig{fig:posanalysis} shows the vehicle trajectories derived from all CAMs captured by one urban and one highway stationary infrastructure station. The assignment of vehicle trajectories to respective streets and driving directions is possible. While trajectories accumulate at lane center lines, there is still quite some variance, blurring the lines between individual lanes. Note that the received CAMs also carry information about localization uncertainty.

\subsection{Suggested Use Cases}
\label{sec:uc}

This paper presents a rudimentary first analysis of the \textit{V2AIX} dataset. Here, we identify potential for further research and usage of the dataset. Some interesting research questions that may be answered based on \textit{V2AIX} and future dataset releases:
\begin{itemize}
    \item How often does one encounter V2X-capable vehicles or other ITS stations in urban/rural/highway environments?
    \item What range does ITS-G5 reach in urban/rural/highway environments in practice?
    \item How accurate is the localization of V2X-capable vehicles in urban/rural/highway environments, compared to the ground-truth extracted from lidar context?
    \item Do the collected real-world messages adhere to the standard in terms of message contents, publication frequency, and more?
    \item To what extent do the collected real-world messages cover and support the breadth of information that is in principle possible to communicate with the standardized message types?
\end{itemize}

\section{Conclusion}
\label{sec:conclusion}

This paper presents \textit{V2AIX}, a novel multi-modal real-world dataset of ETSI ITS messages in public road traffic. With the first iteration of \textit{V2AIX}, we address a notable gap in datasets for automated driving purposes in terms of real-world V2X communication data. Irrespective of the underlying communication technology, the collected ETSI ITS messages are a valuable asset for studying the current market penetration and in-series implementation detail, as well as for developing and testing future C-ITS.

A quantitative and qualitative analysis of the \textit{V2AIX} data has revealed first key insights for mobile and stationary scenarios in urban and highway environments. Using the collected CAM data, we have identified specific V2X-enabled vehicle models and qualitatively investigated their localization accuracy. Additionally, we have found that all required basic state variables such as position and speed are set by the transmitting ITS stations, only some optional properties are missing.

Alongside the dataset, we have open-sourced a contribution to the ROS ecosystem: \texttt{etsi\_its\_messages} provides ROS messages equivalent to the standardized ETSI ITS messages. It also contains a ROS node for bridging ASN.1-encoded V2X payloads to ROS V2X messages and vice-versa. The dataset collection pipeline for \textit{V2AIX} heavily relies on this new helpful tool.

Looking forward, this first release of the \textit{V2AIX} dataset sets a foundation for further research. We have motivated multiple use cases of how the dataset can help to find valuable insights for assessing the current status-quo as well as work on the future generation of V2X and C-ITS. In case of a positive response to \textit{V2AIX}, we plan for a large-scale extension of the dataset with new data: more stationary infrastructure units over a longer period of time as well as mobile measurement drives specifically in regions with active RSU deployment~\cite{eu2022itsmap}. Additionally, we propose to integrate V2X logging into future perception and planning dataset recordings to foster research in cooperative perception and planning.

\section*{Acknowledgements}

This work is accomplished within the projects AIthena, 6GEM and AUTOtech.\textit{agil}. We acknowledge the financial support for the projects by the European Union’s Horizon Europe Research and Innovation Programme under Grant Agreement No 101076754 for AIthena, and the Federal Ministry of Education and Research of Germany~(BMBF) for 6GEM (FKZ~16KISK036K) and \mbox{AUTOtech.\textit{agil}} (FKZ~01IS22088A).

Additionally, we would like to thank Christian Geller, Till Beemelmanns, Jonas Reiher, and Jan Bergmann for their support in collecting and processing data for \textit{V2AIX}.




\bstctlcite{IEEEexample:BSTcontrol}
\bibliographystyle{IEEEtran}
\bibliography{references}

\end{document}